\documentclass[12pt]{article}
\usepackage{epsfig}

\begin{document}

\title{Meson production on complex nuclei by $\pi^-$ \\ with the Crystal Ball 
detector
\footnote{Contribution to the International Workshop on Chiral
Fluctuations in Hadronic Matter, Orsay, France, September 2001.}}

\author{A.~Starostin\footnote{\texttt{starost@physics.ucla.edu}}, B.~M.~K.~Nefkens \\
\emph{UCLA, Los Angeles, CA 90095--1547, USA} \\
H.~M.~Staudenmaier \\
\emph{Universit\"at Karlsruhe, Karlsruhe, Germany 76128} \\
for the Crystal Ball Collaboration
\footnote{Supported in part by US DOE, NSF, NSERC, RMS and VS.}
\footnote{The Crystal Ball Collaboration: 
M.~Clajus, A.~Maru\u{s}i\'c, S.~McDonald, 
B.~M.~K.~Nefkens, N.~Phaisangittisakul, S.~Prakhov, J.~W.~Price,
A.~Starostin, and W.~B.~Tippens, {\em UCLA}, 
D.~Isenhower and M.~Sadler, {\em  ACU}, 
C.~Allgower and H.~Spinka, {\em ANL},
J.~Comfort, K.~Craig, and T.~Ramirez, {\em ASU},
T.~Kycia, {\em BNL},
J.~Peterson, {\em UCo},
W.~Briscoe and A.~Shafi, {\em GWU},
H.~M.~Staudenmaier, {\em UKa},
D.~M.~Manley and J.~Olmsted, {\em KSU},
D.~Peaslee, {\em UMd},
V.~Bekrenev, A.~Koulbardis, N.~Kozlenko, S.~Kruglov, and I.~Lopatin, {\em PNPI},
G.~M.~Huber, G.~J.~Lolos, and Z.~Papandreou, {\em UReg},
I.~Slaus and I.~Supek, {\em RBI},
D.~Grosnick, D.~Koetke, R.~Manweiler, and S.~Stanislaus, {\em ValU}.}}

\maketitle

\begin{abstract}

We report on preliminary results for the production of neutral mesons by 750~MeV/$c$ $\pi^-$ 
on complex nuclear targets (C, Al, Cu) and hydrogen. Our simultaneous measurement of $\pi^0$, 
$2\pi^0$, and $\eta$ final states allows an investigation of the nuclear 
final-state-interaction effects. The data show that nuclear absorption is mainly 
responsible for an observed change in the shape of the $2\pi^0$ invariant mass spectra.
Our preliminary result for the $\sigma \to \gamma \gamma$ branching ratio is 
BR$(\sigma \to \gamma \gamma) < 3.6\times 10^{-3} \times \phi$ at 90\% C.L. 
on a carbon target, where $\phi$ is the fraction of $2\pi^0$ produced via the $\sigma$
intermediate state.

\end{abstract}

\section{Introduction}

The modification of hadron properties in nuclear matter is a major topic in nuclear 
studies. Of special importance is chiral restoration because it is a significant 
milestone on the long path which goes from quark confinement in nuclei at ordinary 
density and temperature to the high density and temperature of the quark--gluon plasma. 
Major changes are expected to occur for the ${\rm f}_0(400 - 1200)$, 
or ``$\sigma$'' state, when it is produced on a nucleus even for low atomic number $A$, 
rather than on a nucleon. The $\sigma$ is the symbol for the correlated $\pi \pi$ system 
in the $J=I=0$ state. It plays an important role in the intermediate range $N N$ 
interaction~\cite{schuck2}. The quantum numbers of the $\sigma$ make it the chiral 
partner of the pion. Under chiral restoration, the $\sigma$ and $\pi$ should have 
degenerate masses, which implies that the $\sigma$ should have a large reduction in mass 
when approaching chiral restoration.

The interest in $\sigma$ medium modifications was fanned in a major way by a 1996 CHAOS 
publication~\cite{chaos1} in which they reported on a measurement of the invariant mass 
spectrum of the $\pi^+\pi^-$ system ($m_{\pi^+\pi^-}$) that was produced by $\pi^+$ 
interactions in complex nuclei. The $m_{\pi^+\pi^-}$ spectrum near $2m_{\pi}$ was reported 
as ``close to zero for $A=2$, and increases dramatically with increasing $A$''. A marked 
$A$--dependent peak at about $2m_{\pi}$ was interpreted by the CHAOS collaboration as 
follows: ``the experimental results indicate that nuclear matter strongly modifies the 
$\pi\pi$ interaction in the $J=I=0$ channel''. The complete $m_{\pi^+\pi^-}$ spectra 
for complex nuclei reported by CHAOS were stunningly similar to theoretical predictions 
of medium modifications made by Schuck {\it et al.}~\cite{schuck1} and Chanfray 
{\it et al.}~\cite{schuck2}. It started a cascade of papers on $\sigma$ medium 
modifications~\cite{theo1,theo2,theo3,chiku,rapp,hat2,vacas} and played a prominent 
role in two recent workshops~\cite{darms,kyoto}. There were no new experimental data 
until the Crystal Ball Collaboration made a careful search for a peak near $2 m_{\pi}$ 
in the more favorable $2\pi^0$ spectra produced by $\pi^-$ interactions. The four photons 
in the $2\pi^0$ final state were detected with the Crystal Ball detector, which has a 
near $4\pi$ acceptance and detects $\pi^0$'s down to zero kinetic energy. There was no 
indication of the CHAOS peak~\cite{2pi0prl}. In the meantime, Bonutti {\it et al.} hinted 
that the CHAOS peak was at least in part an artifact caused by the limited acceptance of 
the CHAOS detector~\cite{chaos2}.

The CHAOS Collaboration has shifted its interest to a change in the shape of the
$m_{\pi^+\pi^-}$ spectra obtained on complex targets~\cite{chaos3}. They proclaim new, 
different evidence for nuclear medium modifications from the shape of the ratio of 
$m_{\pi^+\pi^-}$ for different targets.
However, the difference in the slope of the $m_{\pi^+\pi^-}$ spectra for different $A$ 
can be due to two very different mechanisms. One is the difference in interaction of 
the outgoing particles inside the nucleus. The other one is a generic change in the 
shape to lower $m_{\pi\pi}$ because of $\sigma$ medium modifications.

We present new data on $2\pi^0$, $\pi^0$, and $\eta$ production by $\pi^-$ of 
750~MeV/$c$ that may help in investigating these two possibilities. Specifically, the 
following reactions have been studied:
$A (\pi^-,\pi^0 \pi^0 ) X^0$,
$A (\pi^-,\eta ) X^0$ (via the $\eta \to \gamma \gamma$ decay mode), and
$A (\pi^-,\pi^0 ) X^0$,
where $A$ is H, C, Al, or Cu. We will show experimental similarities and 
differences that have been found for the reactions above. Our principal focus will be on 
the reaction $A (\pi^-,\pi^0 \pi^0 ) X$. 
\begin{figure}
\centerline{\psfig{file=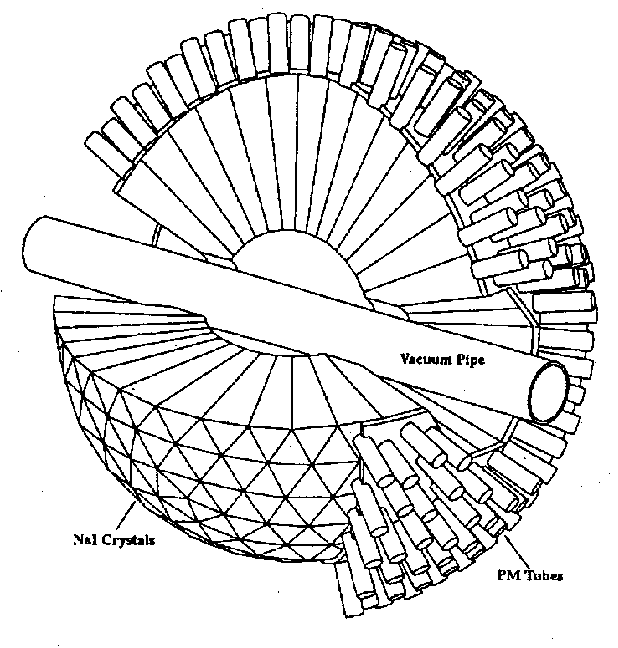,width=0.53\textwidth}
            \psfig{file=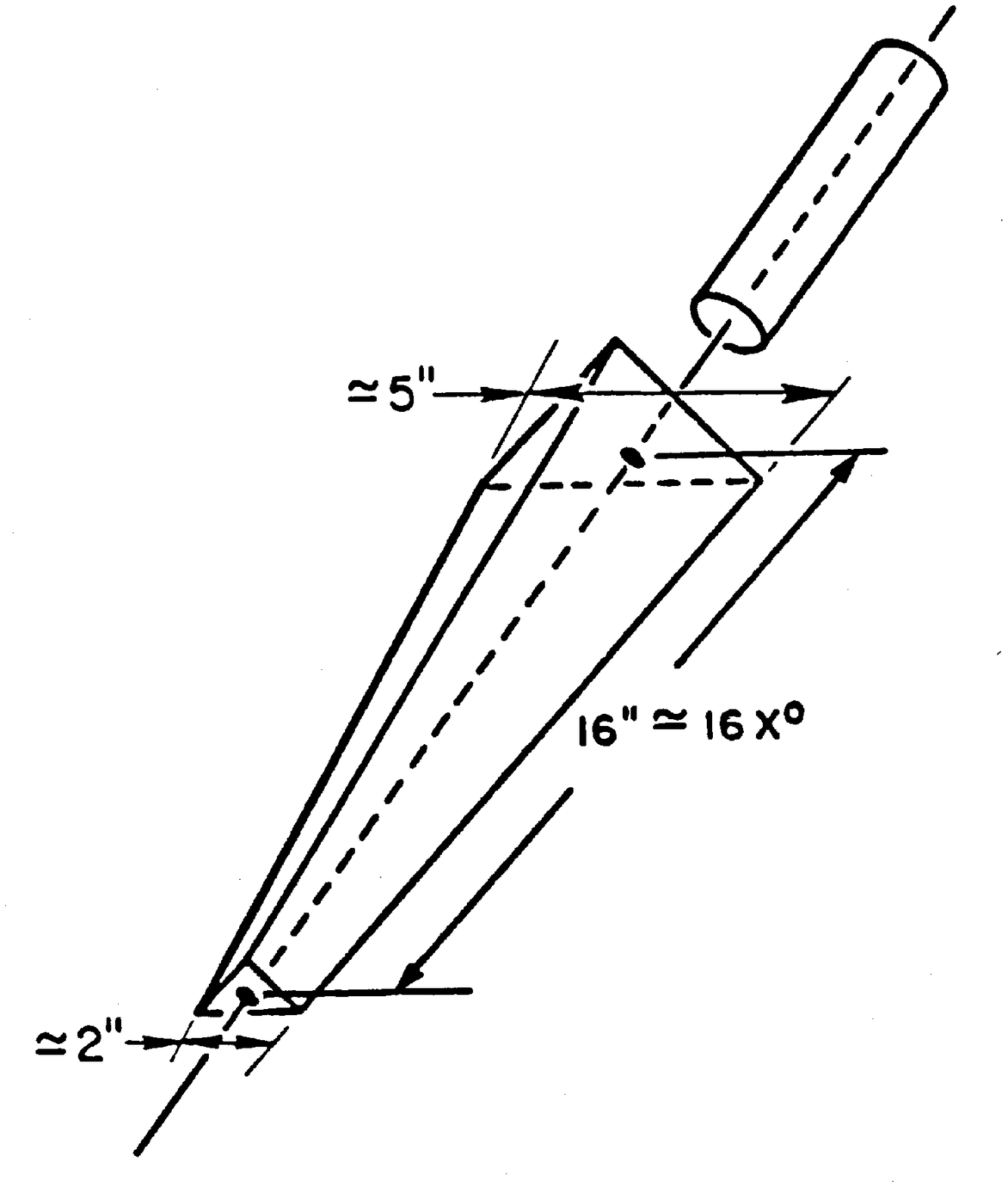,width=0.47\textwidth}}
\caption{On the left is a schematic diagram of the Crystal Ball detector. 
On the right are the dimensions of a typical individual crystal.}
\label{cball}
\end{figure}

\section{Experimental method}

Photons from $\pi^0$ and $\eta$ decays were detected with the Crystal 
Ball (CB) multiphoton spectrometer (see Fig.~\ref{cball}).
It is constructed of 672 optically isolated NaI(Tl) crystals that cover 93\% of $ 4\pi$ 
sterad. Electromagnetic showers in the spectrometer are measured with an
energy resolution $\sigma_{s.d.}^E/ E \sim 1.7\%/(E\ ({\rm GeV}))^{0.4}$;
the angular resolution for photon showers at energies 50--500~MeV
is $\sigma_{s.d.}^\theta = 2^\circ$--$3^\circ$ in the polar
angle and $\sigma_{s.d.}^\phi = 2 ^\circ/\sin\theta$ in the azimuthal angle.
Here $\sigma_{s.d.}$
is the standard deviation for a Gaussian distribution. 
The neutral final states were distinguished from charged ones by
a veto barrel made of four plastic scintillation counters 
that covered the active volume of the CB. Most of our results presented here were 
obtained for the neutral final states. The charged final states were used to
calculate the experimental corrections due to photon 
conversion and nuclei breakup.

The experiment was performed in the C6 beam line of the Brookhaven National Laboratory AGS.
The centroid of the beam momentum is known to better than 1\%. We used three solid
targets: C (3.44~g/cm$^2$), Al (1.69~g/cm$^2$),
and Cu (1.51~g/cm$^2$). A 10~cm--long liquid hydrogen target (LH$_2$) was used for 
the hydrogen data presented here.

The reaction $A (\pi^-,\pi^0 \pi^0 ) X^0$ was identified from events with four clusters.
Two-cluster events were used to select the reactions $A (\pi^-,\eta ) X^0$ and 
$A (\pi^-,\pi^0 ) X^0$. See Refs.~\cite{2pi0prl,etalambda} for more details on the 
experimental apparatus, data analysis, and the Monte Carlo simulation for acceptance 
calculations. In order to facilitate comparisons for different targets, we calculated an 
``effective missing mass'' for each nuclear target, given by
\begin{equation}
M_{\rm target}(A) = M(A,Z) - M(A-1,Z-1),
\end{equation} 
where $M(A,Z)$ is the mass of the target nucleus, $M(A-1,Z-1)$ is the mass of the residual 
nucleus, and $Z$ is the atomic number of the target nucleus. 
Missing mass spectra for hydrogen as well as the effective missing mass spectra 
for nuclear targets were calculated for all three reactions assuming 
``quasifree'' production.

\section{Study of $\pi^0\pi^0$ production}
\label{sec2pi0}

\begin{figure}
\centerline{
\parbox{0.5\textwidth}{
\epsfig{file=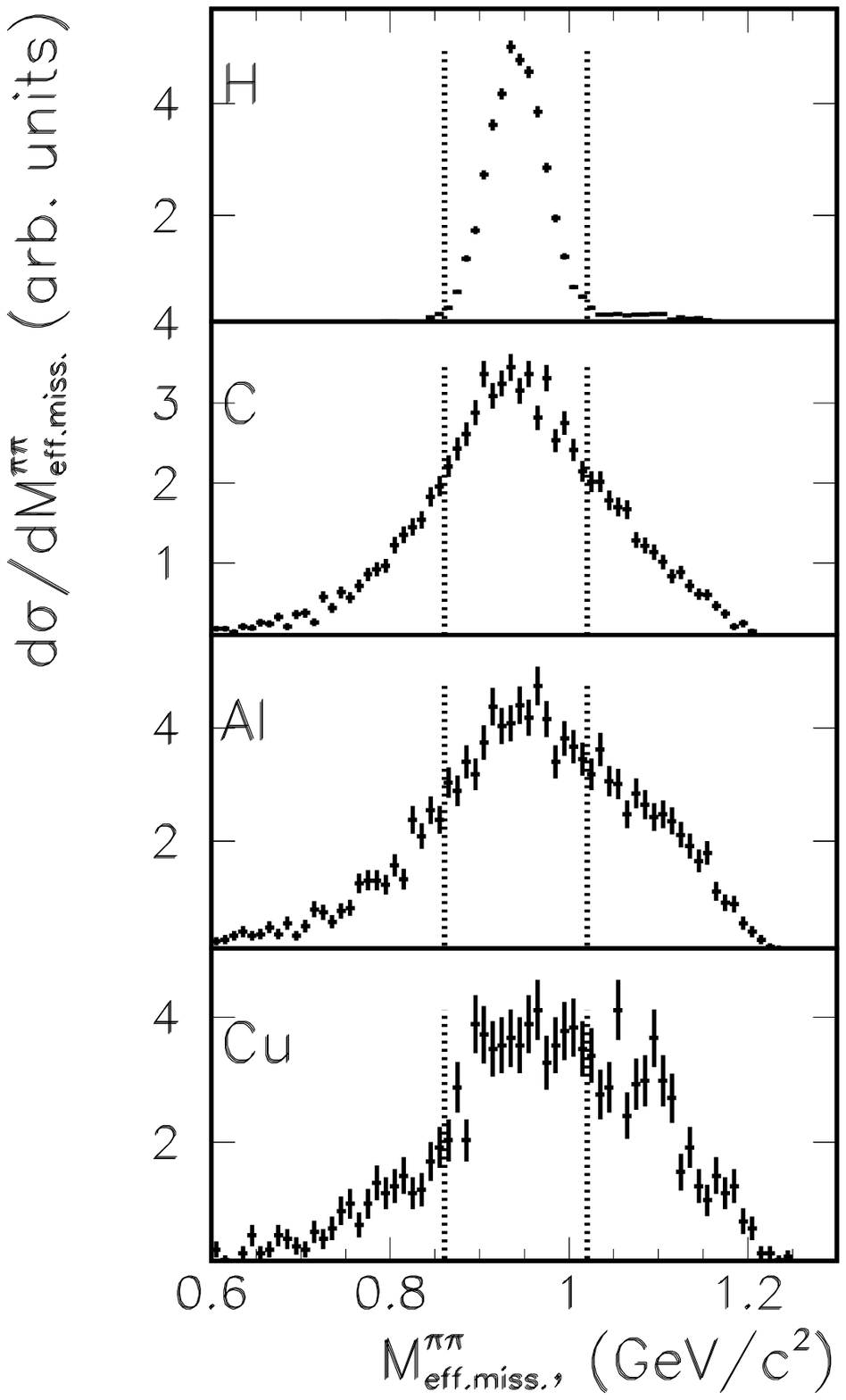,width=0.5\textwidth}}
\parbox{0.5\textwidth}{
\epsfig{file=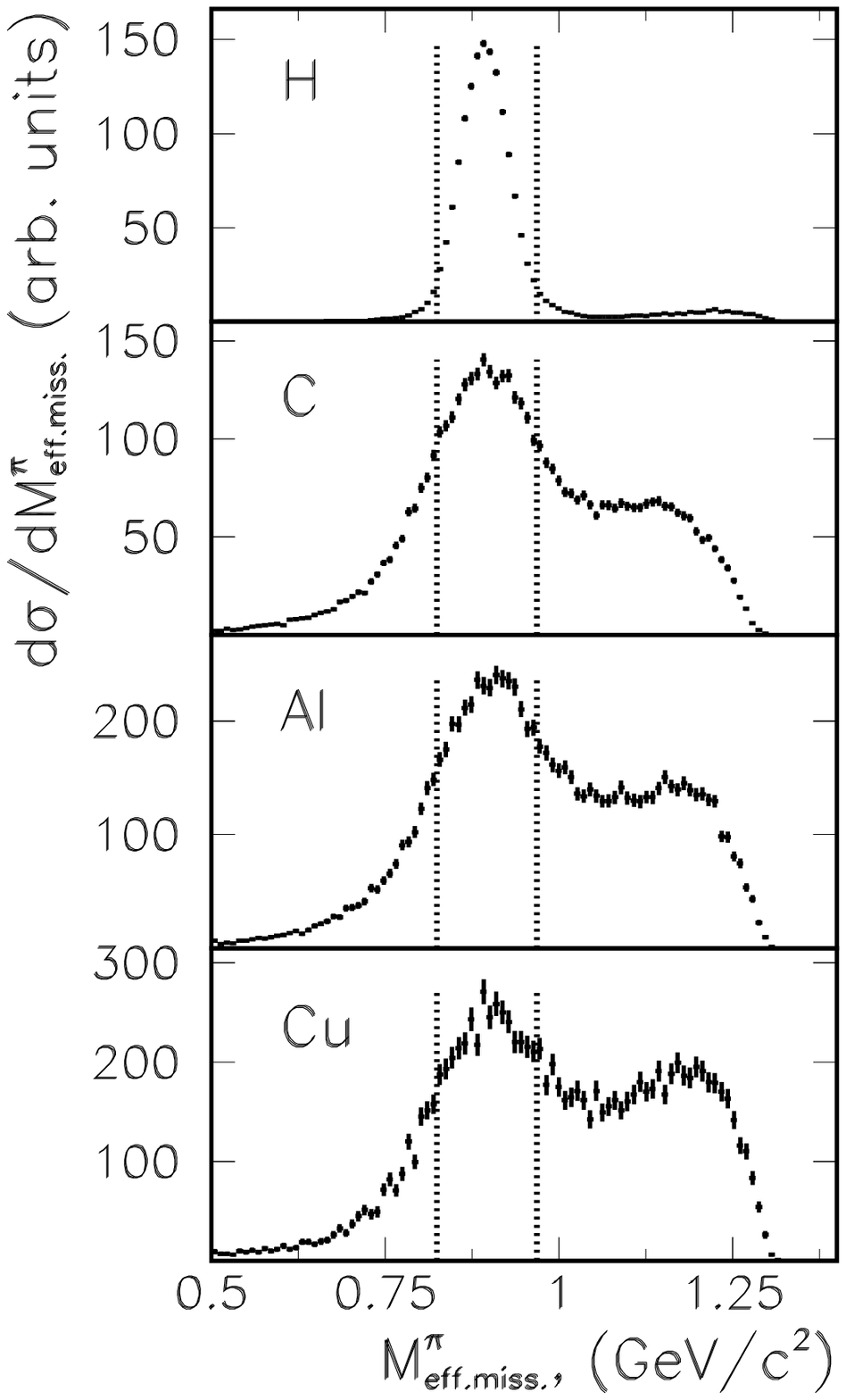,width=0.5\textwidth}}}
\caption{{\bf Left:} Missing-mass distribution for $2\pi^0$ events from hydrogen and 
effective-missing-mass distributions on C, Al, and Cu. 
{\bf Right:} Missing-mass distribution for single $\pi^0$ events from hydrogen and 
effective-missing-mass distributions from C, Al, and Cu. The dashed lines show
our cuts to be applied to the effective missing mass.}
\label{fig3}
\end{figure}

\begin{figure}
\centerline{
\parbox{0.5\textwidth}{
\epsfig{file=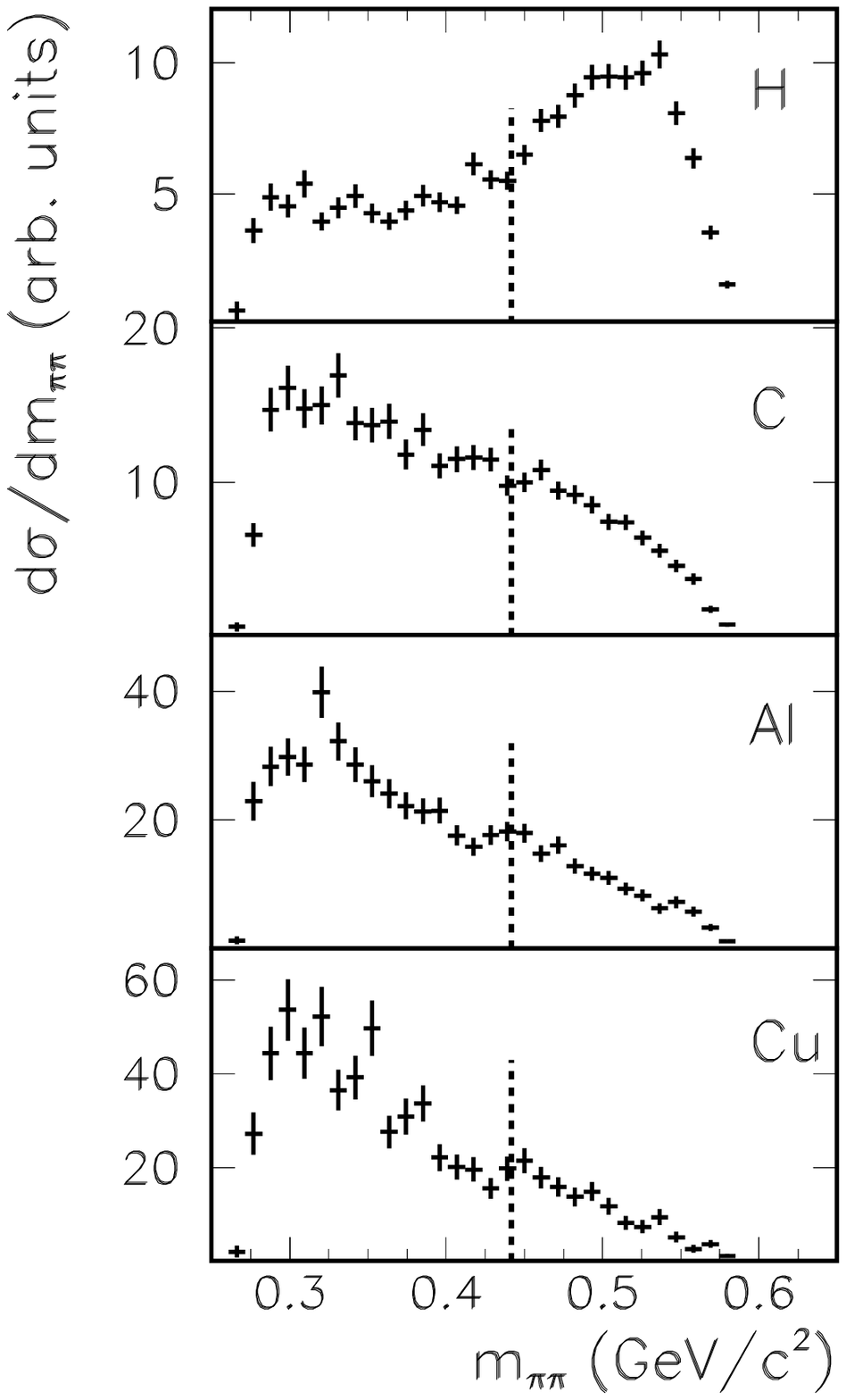,width=0.5\textwidth}}
\parbox{0.5\textwidth}{
\epsfig{file=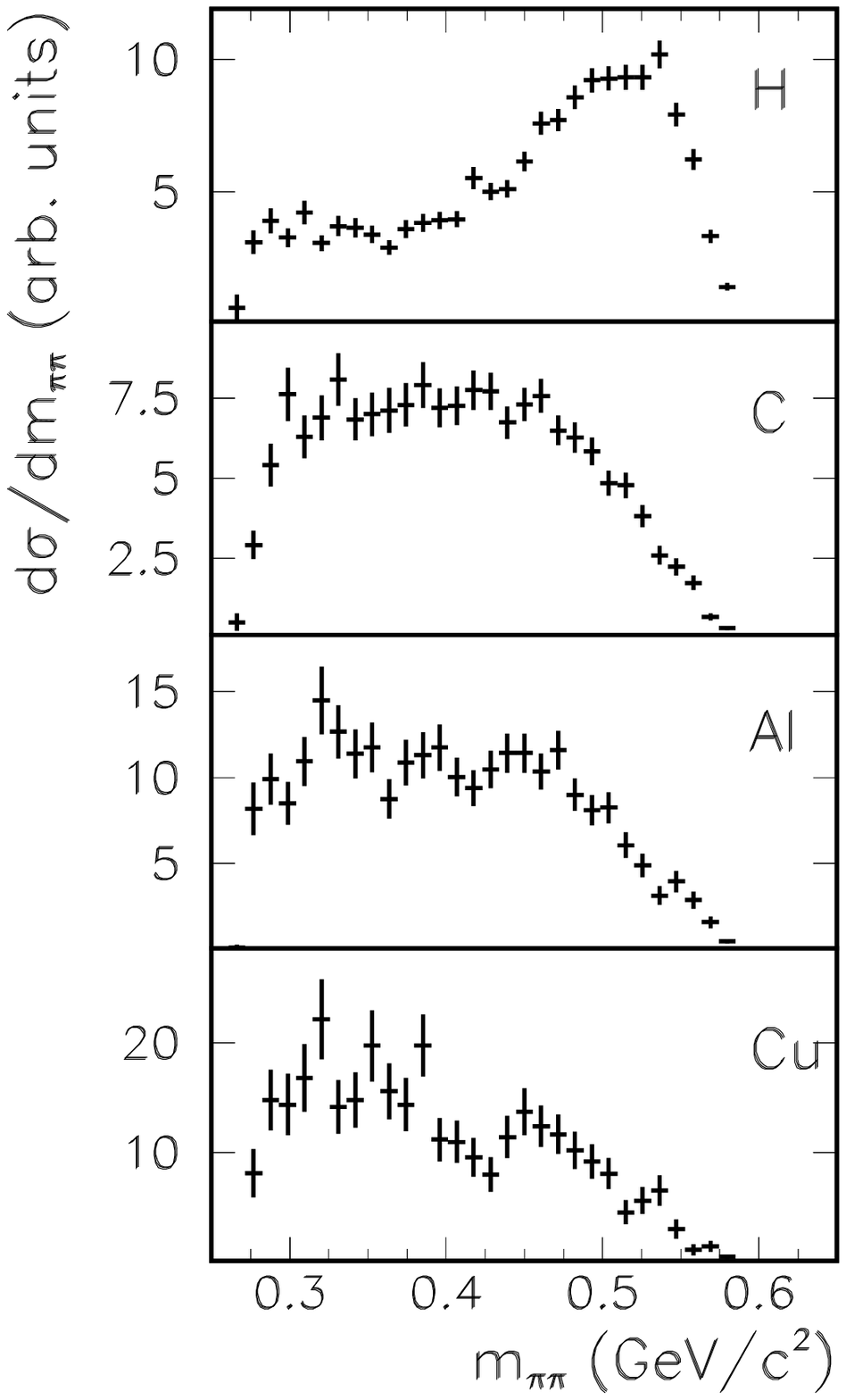,width=0.5\textwidth}}}
\caption{
{\bf Left:} Preliminary $2\pi^0$ invariant-mass distributions obtained for H, C, Al, and Cu 
without the cuts on the effective
missing mass. The spectra have been corrected for the small variation in the CB acceptance. 
Only statistical uncertainties are shown. The dashed lines indicate the
``low'' and ``high'' invariant-mass regions used for the calculation of $\sigma_{int}$. 
{\bf Right:} Preliminary $2\pi^0$ invariant-mass distributions obtained with the 
effective-missing-mass cuts indicated in Fig.~\ref{fig3}.
The spectra have been corrected for the CB acceptance. 
The same correction was used as in the left figure. Only statistical uncertainties are shown.
}
\label{fig5}
\end{figure}

The missing mass distributions of $2\pi^0$ events obtained with the 
LH$_2$ target as well as the spectra of the effective missing mass for
different nuclear targets are presented in Fig.~\ref{fig3}. A sharp
neutron peak with width $\sigma_{s.d.} \approx 35$~MeV is seen in case of 
the hydrogen target. For the solid targets this peak
broadens substantially. Some of the broadening is explained by the Fermi
momentum of the target nucleons. Our Monte Carlo calculations indicate
that the broadening is mainly due to nuclear
effects such as $\pi^0$ rescattering inside the nucleus. 
Our calculations of the pion rescattering and absorption used
the model developed in Ref.~\cite{mpath}. 
To select $2\pi^{0}$ events with minimum rescattering distortions, we applied the
effective-missing-mass cuts indicated in Fig.~\ref{fig3} by the two vertical dashed lines. 
The resulting $2\pi^{0}$ invariant mass distributions,
both with and without the missing mass cut, are given in Fig.~\ref{fig5}. 
All spectra are compatible with a smooth distribution; {\it i.e.}, there is no sharp peak.
The broad maximum at $m_{\pi^0\pi^0} \approx 510$~MeV obtained for the
hydrogen target disappears for carbon and other nuclear targets. 
Our invariant-mass distributions for C, Al, and Cu 
targets obtained without the cuts on the effective missing mass show broad maxima for 
$2\pi^0$ masses below 400~MeV. The distributions obtained for C and Al become almost flat 
between 300~MeV and 480~MeV after the application of the effective-missing-mass cuts.

\begin{figure}
\centerline{
\parbox{0.5\textwidth}{
\epsfig{file=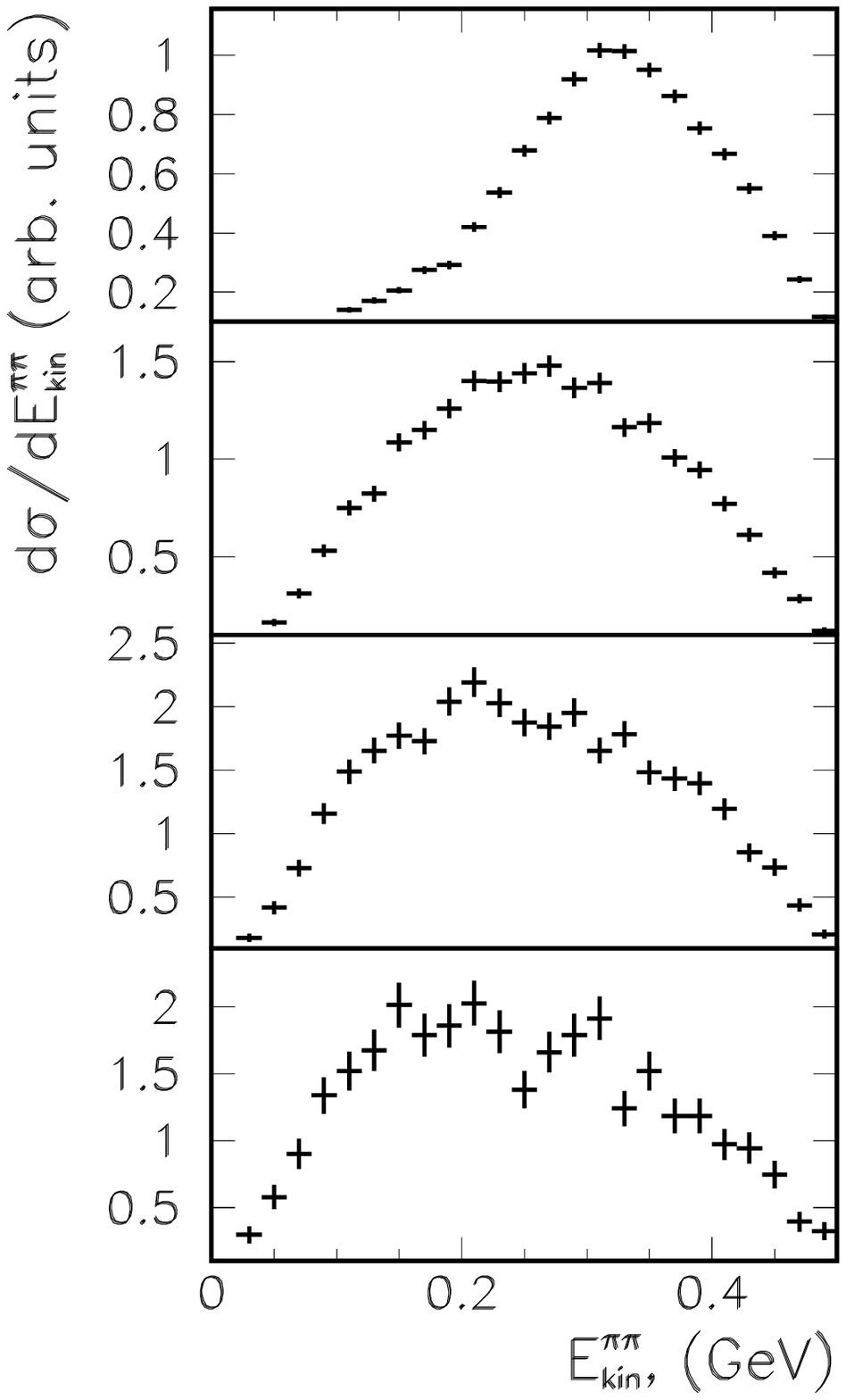,width=0.5\textwidth}}
\hspace{0.1\textwidth}
\parbox{0.5\textwidth}{
\epsfig{file=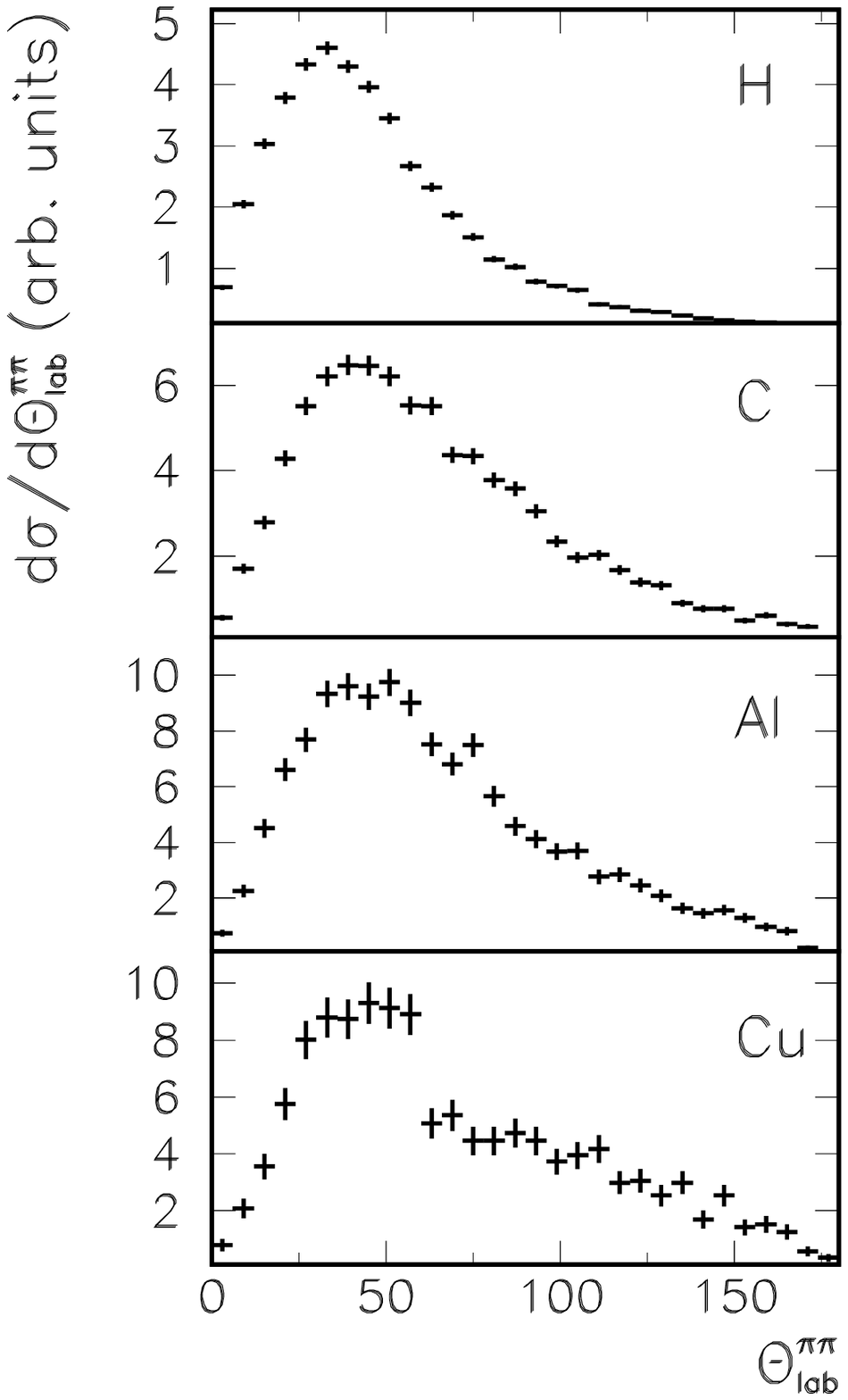,width=0.5\textwidth}}}
\caption{{\bf Left:} Kinetic energy distribution of the $\pi^0\pi^0$ system.
{\bf Right:} Angular distribution in the laboratory of the $\pi^0\pi^0$ system.
Neither distribution is corrected for the small variation in the Crystal Ball acceptance. 
The acceptance is a smooth function of the kinetic energy and the lab angle. 
Only statistical uncertainties are shown.}
\label{fig6}
\end{figure}

The kinetic-energy spectra
of the $\pi^{0}\pi^0$ system are shown in Fig.~\ref{fig6} for our targets. 
The spectra differ markedly for various targets.
The angular distributions for the $2\pi^0$ system in the laboratory 
for different targets are shown in Fig.~\ref{fig6}. The shapes of the angular 
distributions have similar features such as the broad maximum at about $25^\circ$ 
for all targets including hydrogen. The minor increase in the broadness of 
the maxima as a function of $A$ can be explained by the Fermi momentum of the target 
protons. Neither the distribution of the kinetic energy nor the angular distributions were 
corrected for the Crystal Ball acceptance. The acceptance is a rather flat and smooth 
function of the kinetic energy and the lab angle.

To probe the net effect of the nuclear medium, we have investigated the cross sections
as a function of $A$. To trace the disappearance of the high mass 
peak near 500~MeV in the invariant mass spectra, we have divided all distributions 
in Fig.~\ref{fig5} as indicated into a low--mass region (between 270~MeV and 445~MeV) and a 
high--mass region (above 445~MeV). The angular $\Theta_{lab}$ distributions were 
used to calculate integrated cross sections, $\sigma_{int}^{low}$ and 
$\sigma_{int}^{high}$, for both mass regions as well as the total integrated cross 
section $\sigma_{int}$. Assuming that the nuclear dependence of the 
cross section is proportional to the geometrical factor $\lambda_{geom}$, where
\begin{equation}
\lambda_{geom} = A^{2/3}\frac{Z}{A} = Z A^{-1/3},
\label{s_int}
\end{equation} 
\noindent we normalized our cross sections to $\lambda_{geom}$ and to 
the hydrogen cross section, $\sigma_{int}$(H), combining them into the ratio $R$:
\begin{equation}
R = \frac{\sigma(\pi^- A \to \pi^0 \pi^0 X)}{\sigma(\pi^- p \to \pi^0 \pi^0 n)} 
\times \frac{A^{1/3}}{Z}.
\label{rati}
\end{equation} 
\begin{figure}
\centerline{
\parbox{0.5\textwidth}{
\epsfig{file=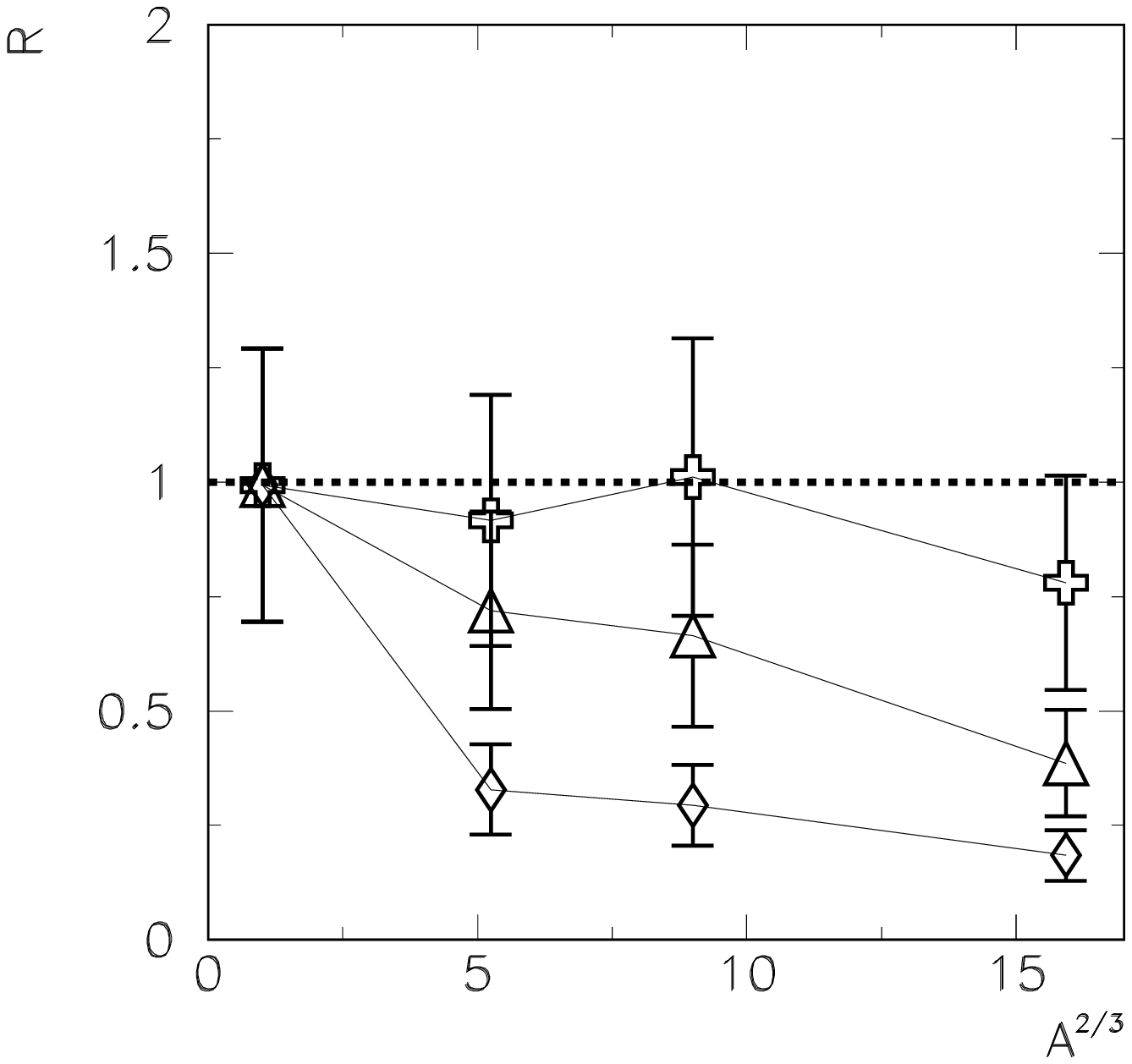,width=0.5\textwidth}}
\parbox{0.5\textwidth}{
\epsfig{file=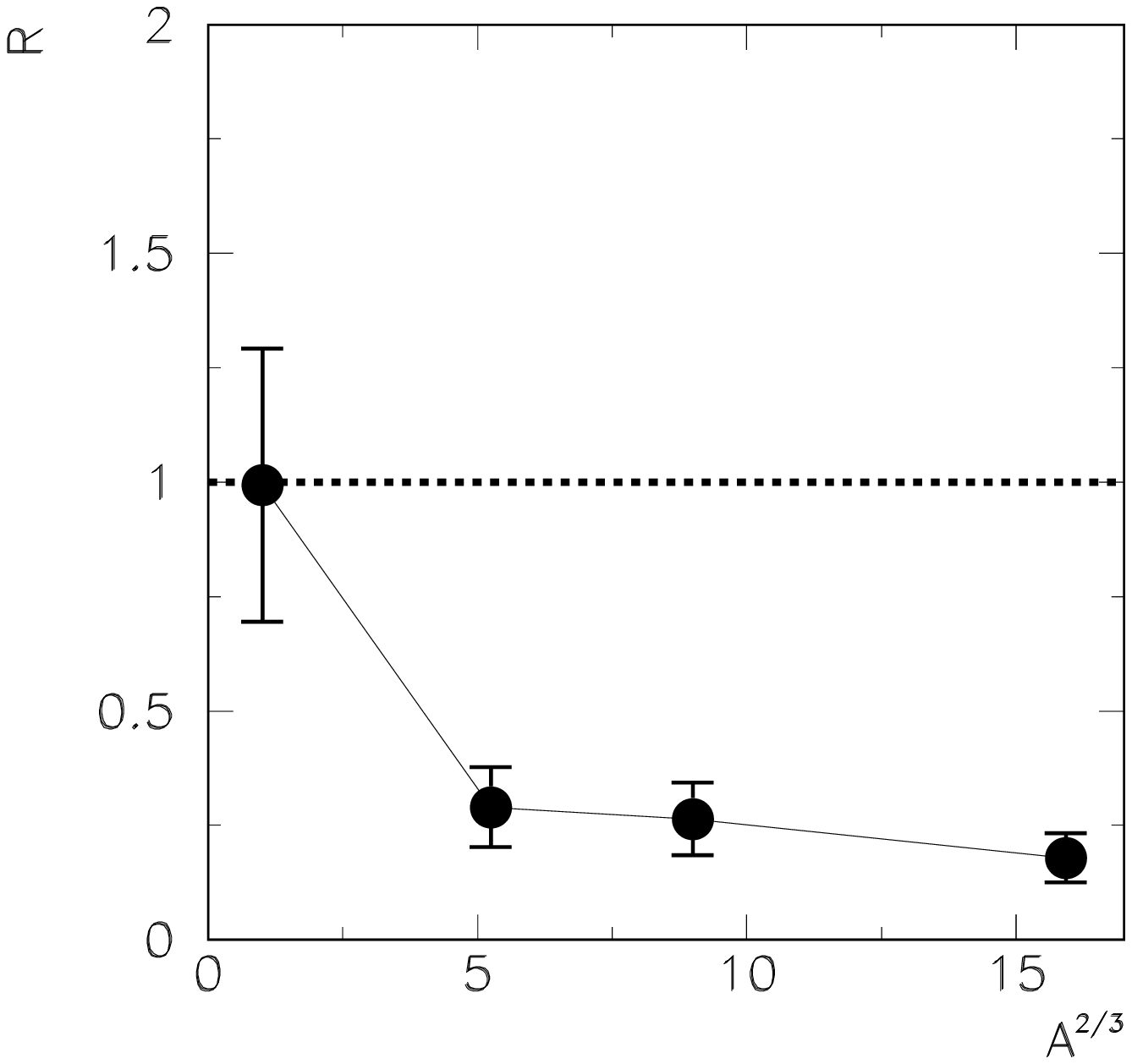,width=0.5\textwidth}}}
\caption{
{\bf Left:} Ratios of normalized $2\pi^0$ total cross sections as a function of $A^{2/3}$.
The result for $\sigma_{int}^{low}$ is shown by crosses, 
$\sigma_{int}^{high}$ by diamonds, and $\sigma_{int}^{total}$ by triangles.
{\bf Right:} Ratios of normalized $\eta$ total cross sections as a function of $A^{2/3}$.
Statistical and 30\% systematical uncertainties have been added in quadrature for both 
distributions. The systematical uncertainty originates mainly from the loss of the good 
events that appear in the charged mode due to photon conversion in the targets and/or 
charged products of nuclear reactions hitting the veto-barrel counters. The results are 
preliminary.} 
\label{fig7}
\end{figure}
The results for $\sigma_{int}^{low}$, 
$\sigma_{int}^{high}$, and $\sigma_{int}^{total}$ for C, Al, and Cu are shown in 
Fig.~\ref{fig7} as a function of $A^{2/3}$. It can be seen that $\sigma_{int}^{low}$ remains
rather constant and close to the expectation based on Eq.~\ref{s_int}, which is shown in 
Fig.~\ref{fig7} by a straight line. In contrast, the $\sigma_{int}^{high}$ cross section
falls off very rapidly from hydrogen to carbon; {\it i.e.}, the maximum at about 510~MeV
seen in the $\pi^0\pi^0$ invariant mass spectra (see Fig.~\ref{fig5})
is absorbed rather than shifted toward lower masses. For higher $A$'s 
the $\sigma_{int}^{high}$ remains nearly constant. In summary, Fig.~\ref{fig7}
shows that mainly high--mass $2\pi^0$ events are absorbed by nuclear medium 
effects, whereas low mass events are less effected. In the case of $\pi^0 \Delta^0$ 
production, the events may be absorbed via the $\Delta N \to N N$ mechanism.

\section{Study of $\eta$ production}
\label{sect_eta}

Though $\pi^{0}$ and $\eta$ mesons are similar in some aspects {\it e.g.}, 
they both are pseudoscalar mesons with $J^{PC} = 0^{-+}$, they also have relevant differences 
in isospin and in quark composition. It is therefore very attractive to compare
the behavior of $\pi^{0}$ and $\eta$ mesons in nuclear matter. The theoretical interest 
has been discussed by Y.~S.~Golubeva {\it et al.}~\cite{golub}.

\begin{figure}
\epsfig{file=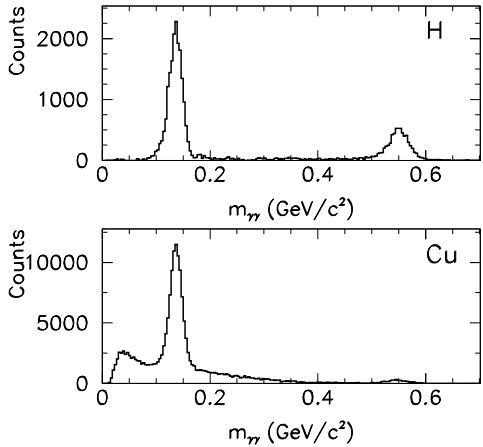,width=0.5\textwidth}
\caption{
Invariant-mass distributions of two photons for H (upper) and Cu (lower) targets after 
subtraction of the empty-target contribution, which is only a few percent.
}
\label{fig1}
\end{figure}
All $\eta$ results in this section are preliminary. The data were obtained from the 
$\eta \to 2\gamma$ decay mode. Figure \ref{fig1} shows the $2\gamma$ invariant mass spectra 
for the H and Cu targets. The first peak in the spectrum is due to $\pi^0$ 
(width $\sigma_{s.d.} \approx 11.5$~MeV), and the second 
peak is due to $\eta$ ($\sigma_{s.d.} \approx 18.0$~MeV). The background under the 
$\pi^0$ and $\eta$ peaks is associated mainly with photon/neutron misidentification. 
In the case of copper, some additional background arises from products of nuclei breakup 
and recoil interactions. 
\begin{figure}
\centerline{
\parbox{0.5\textwidth}{
\epsfig{file=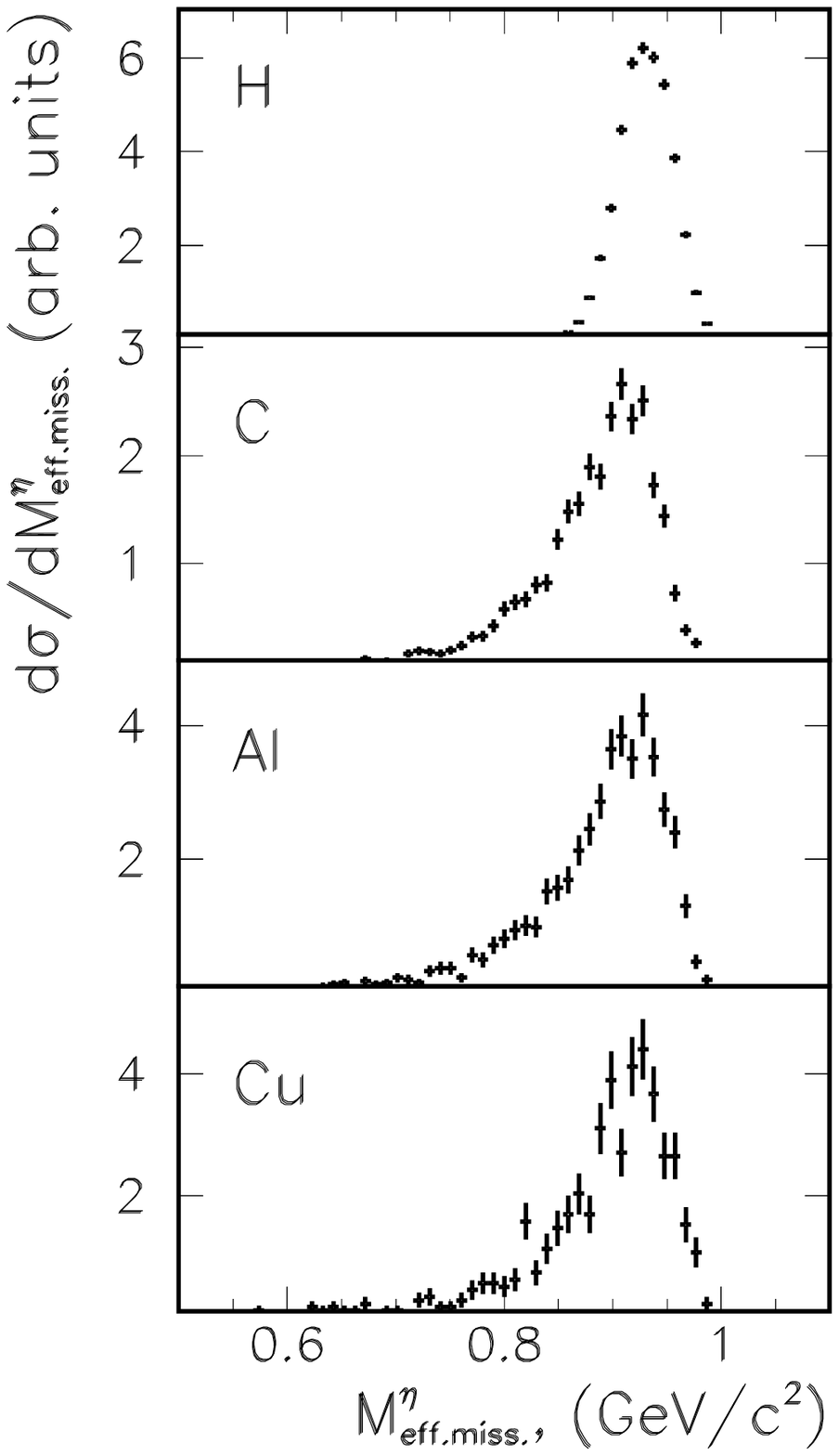,width=0.5\textwidth}}
\parbox{0.5\textwidth}{
\epsfig{file=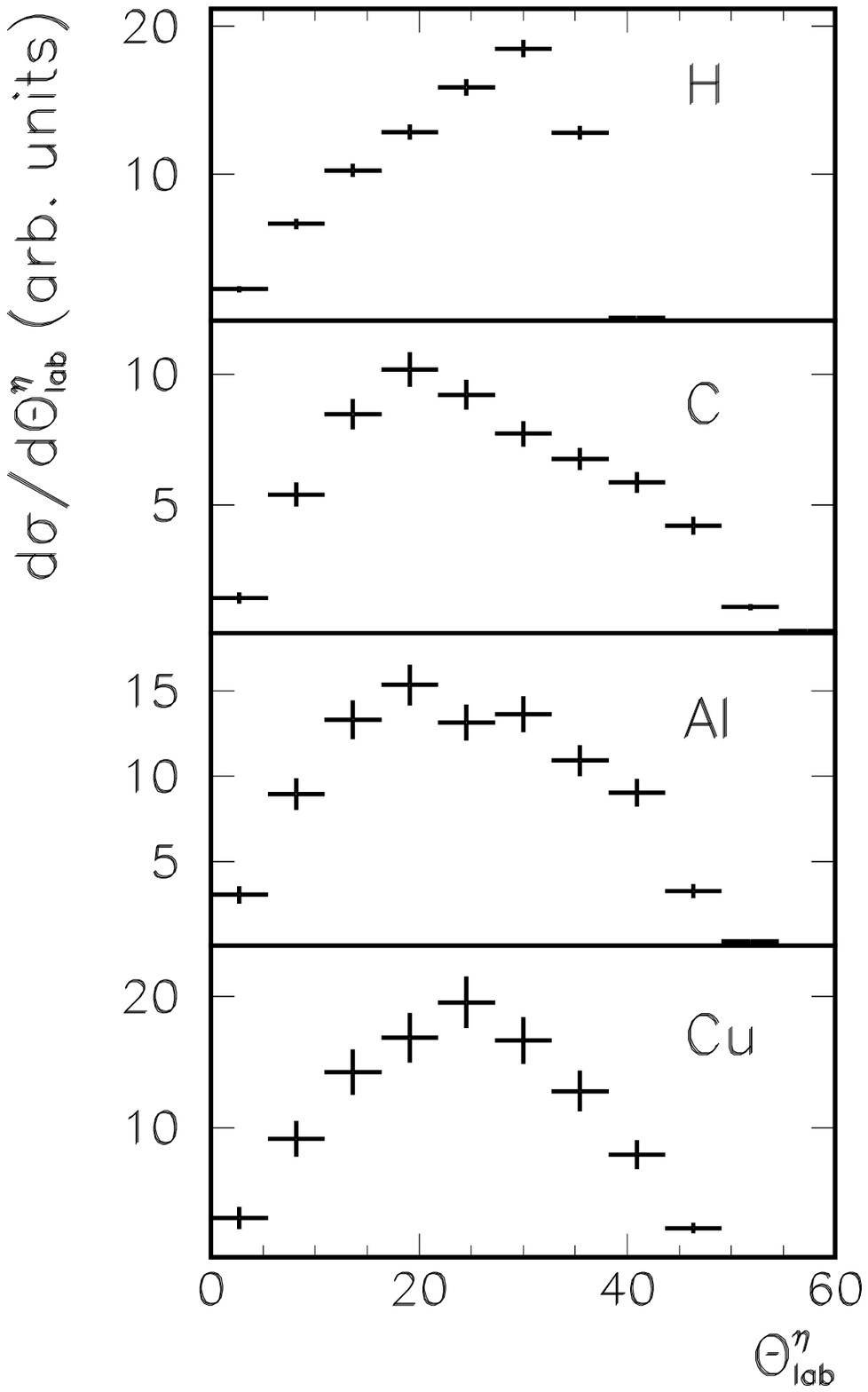,width=0.5\textwidth}}}
\caption{
{\bf Left:} Missing-mass distributions for $\eta \to 2\gamma$ events on hydrogen and 
effective-missing-mass distributions for C, Al, and Cu.
{\bf Right:} Angular distribution of $\eta \to 2\gamma$ events.
The distributions have not been corrected for the Crystal Ball acceptance. 
Only statistical uncertainties are shown.
}
\label{fig8}
\end{figure}

The missing-mass and the effective-missing-mass 
distributions for $\eta$ production are shown in Fig.~\ref{fig8}. 
A nearly background-free neutron peak can be seen for hydrogen. 
The widening of the peaks towards lower values of the effective missing mass for carbon and 
other nuclear targets is less dramatic than for the $\pi^0 \pi^0$ case. It can be explained 
mainly by the nucleon Fermi momentum. It indicates that $\eta$'s are less affected by 
rescattering than pions. A similar behavior is observed in the
$\eta$ angular distribution in the laboratory system (see Fig.~\ref{fig8}). The $\eta$ 
angular distribution in the lab for hydrogen at this momentum is limited to about $40^\circ$ 
by kinematics. It peaks at about $38^\circ$ due to the Jacobian. For the solid targets, 
the angular distributions slightly exceed the two--body kinematical limit because of the 
Fermi momentum. At 750~MeV/$c$, the CB acceptance for this reaction is smooth and uniform.

By integrating the angular distributions, we obtained 
integrated cross sections $\sigma_{int}$ for the H, C, Al, and Cu targets.
$\sigma_{int}$ was normalized to $\lambda_{geom}$ and to the hydrogen data. The results 
shown in Fig.~\ref{fig7} indicate that the $\eta$ cross section falls from H to C, and it is
lower than the expectation based on the geometrical factor (see Eq.~\ref{s_int}). 
It confirms the large $\eta N$ absorption cross 
section observed in earlier $\eta$ photoproduction experiments~\cite{kruge}.
From a comparison of the ratios in Fig.~\ref{fig7}, one may conclude that the behavior of 
$\sigma_{int}(\eta)$ is remarkably similar to that obtained for the 
$\sigma_{int}^{high}(\pi^0\pi^0)$ ratio.

We summarize the results of this section as follows:
\begin{enumerate}
\item The shape of the $\eta$ missing mass is little influenced by nuclear media 
(see Fig.~\ref{fig8}).
\item A strong $\eta$ absorption in nuclear targets was observed (see Fig.~\ref{fig7}).
\end{enumerate}

\section{Study of single $\pi^0$ production}
\begin{figure}
\centerline{
\parbox{0.5\textwidth}{
\epsfig{file=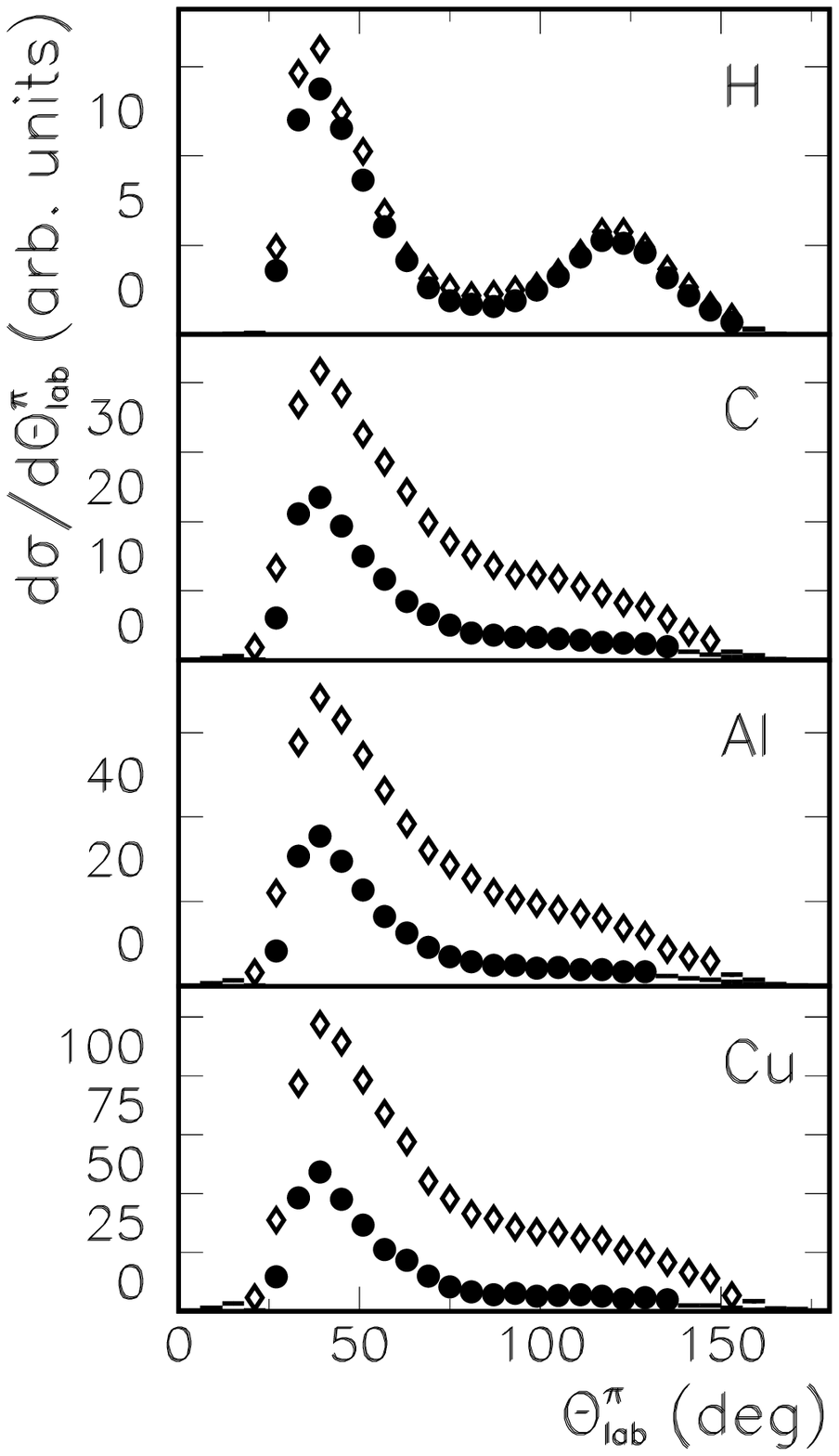,width=0.5\textwidth}}
\parbox{0.5\textwidth}{
\epsfig{file=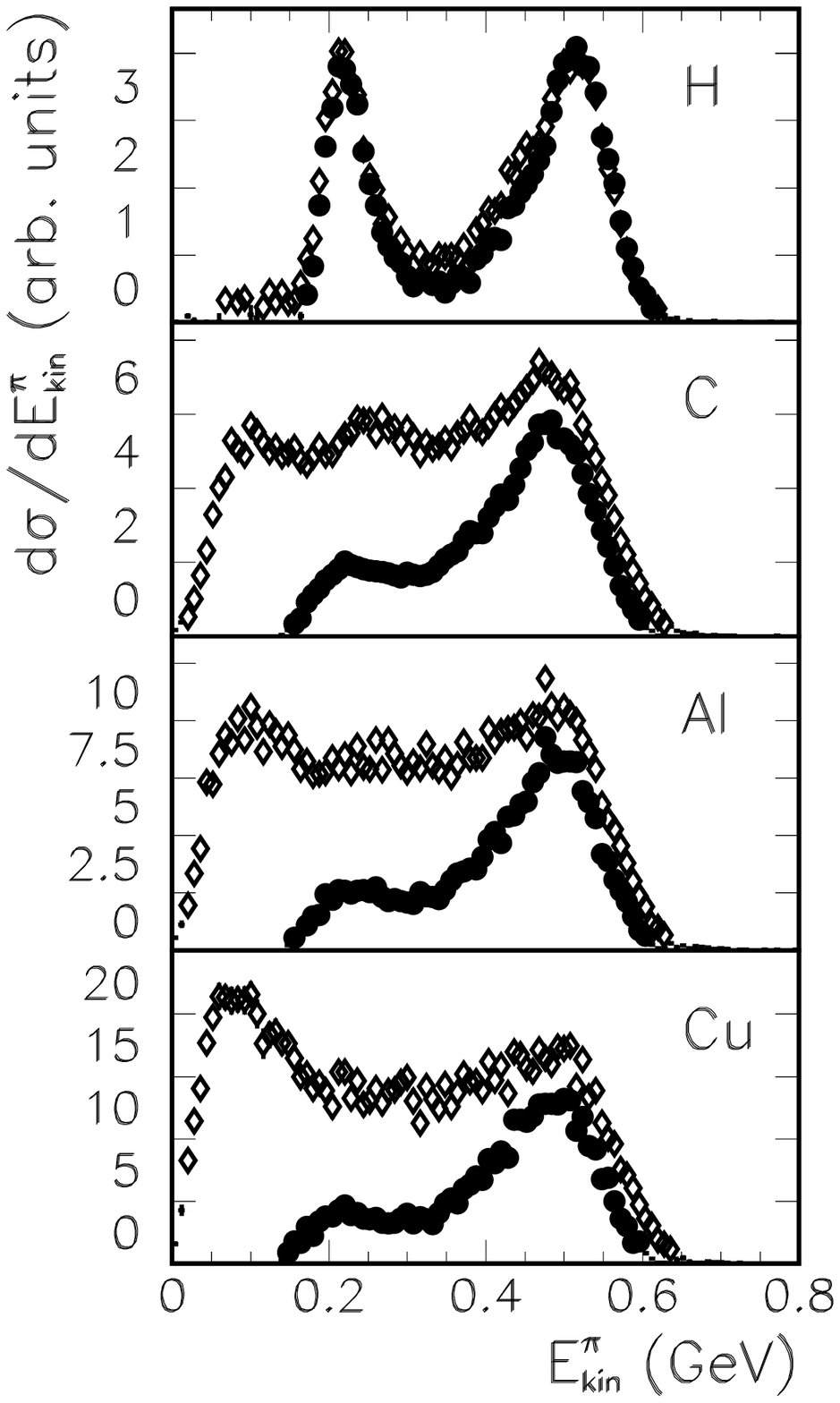,width=0.5\textwidth}}}
\caption{
{\bf Left:} Preliminary results for the angular distributions of single-$\pi^0$ events.
{\bf Right:} Preliminary results for the distributions of the $\pi^0$ kinetic energy for 
different targets. On both plots the open diamonds show the distributions without the 
effective-missing-mass cuts, and the solid circles show the distributions with the cuts. 
None of the distributions have been corrected for the variation in the Crystal Ball 
acceptance. Only statistical uncertainties are shown.
}
\label{fig10}
\end{figure}

Single $\pi^0$ production is the predominant reaction among the three we have studied. 
In the case of hydrogen, the charge-exchange reaction (CEX) is
well understood in terms of conventional $\pi N$ phase-shift analyses. For nuclear
targets, however, CEX is accompanied by strong secondary reactions that lead to a large 
fraction of single $\pi^0$ produced in reactions other than direct CEX. One of the 
contributing reactions is $\Delta^0$ production via $\pi^- p \to \pi^0 \Delta^0$. The 
$\Delta^0$ can be rescattered or absorbed on other nucleons, {\it e.g.}, 
$\Delta N \to N N$, and the reaction ends with a single $\pi^0$. 
In that case, the missing mass would be equal to the mass of the $\Delta^0$.
Figure~\ref{fig3} shows the single-$\pi^0$ missing mass and the effective-missing-mass 
spectra for neutral two-cluster events. Compared to hydrogen, a substantial broadening of 
the ``quasi-elastic'' peak can be seen in Fig.~\ref{fig3} for all nuclear targets. This 
broadening reflects a large probability for the $\pi^0$ to rescatter inside the nuclei. 
In contrast to the hydrogen data, the C, Al, and Cu effective-missing-mass data show a 
second maximum at masses of about 1200~MeV. Such maxima can be partially attributed to 
$\pi^0 \Delta^0$ production followed by $\Delta^0$ absorption in nuclei. This observation 
is in agreement with the conclusions of Sec.~\ref{sec2pi0} regarding the absorption of 
$2\pi^0$ events in nuclei. 

For some applications it may be desirable to separate partly the secondary processes from 
the ``quasi-elastic'' CEX events by applying the cuts on the effective missing mass. 
The influence of those to-some-extent arbitrary cuts in the angular
distribution and the pions kinetic energy is shown in Fig.~\ref{fig10}. 
The distributions are not corrected for the 
Crystal Ball acceptance. The CB acceptance is smooth and an uniform function between 
30$^\circ$ and 150$^\circ$. The acceptance for $\pi^0$ drops very rapidly below 
30$^\circ$ and above 150$^\circ$, due to the entrance and exit tunnels. In the angular 
distribution obtained for hydrogen, a clear backward peak is seen. The peak disappears 
for the other targets. A similar behavior for single $\pi^0$ production 
on carbon was found by R.~J.~Peterson {\it et al.}~\cite{jerry}. For a two--body kinematic 
process, the backward peak in 
the $\pi^0$ angular distribution corresponds to pions with kinetic energy of about 200~MeV. 
The peak is clearly seen in the single-$\pi^0$ kinetic-energy distribution for hydrogen 
(see Fig.~\ref{fig10}). The peak disappears for carbon and other nuclear targets, which 
indicates a large probability for the absorption of $\pi^0$'s with kinetic energy of about 
200~MeV. A possible mechanism for such absorption is the formation of a 
$\Delta$ inside nuclei in $\pi^0 N$ interactions followed by $\Delta$ absorption.

\section{Upper limit for BR$(\sigma \to 2\gamma)$ decay in nuclear media}

The diphoton decay $\sigma \to 2 \gamma$ in hadronic matter has been evaluated in 
Ref.~\cite{chiku} from optimized perturbation theory. The $\sigma \to 2\gamma$ decay is 
interesting because the photons are not affected by final-state interactions. Despite the 
low branching ratio, it might be possible to see the decay experimentally if there is a 
prominent $\sigma$ enhancement at low invariant mass in the nuclear medium. The main 
background is the thermal annihilation of two pions, $\pi^+ \pi^- \to 2 \gamma$.
\begin{figure}
\centerline{
\parbox{0.5\textwidth}{
\epsfig{file=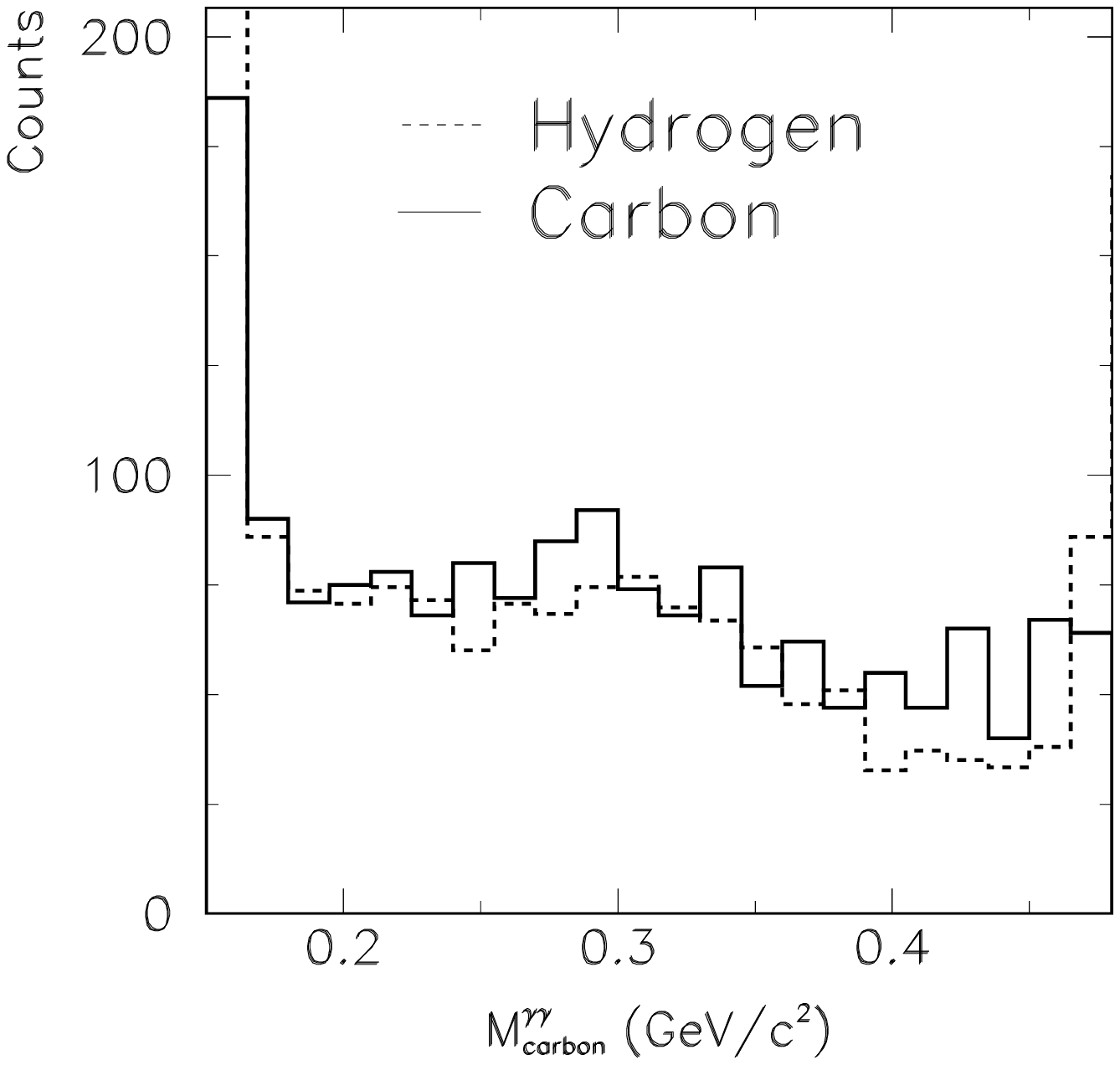,width=0.5\textwidth}}
\parbox{0.5\textwidth}{
\epsfig{file=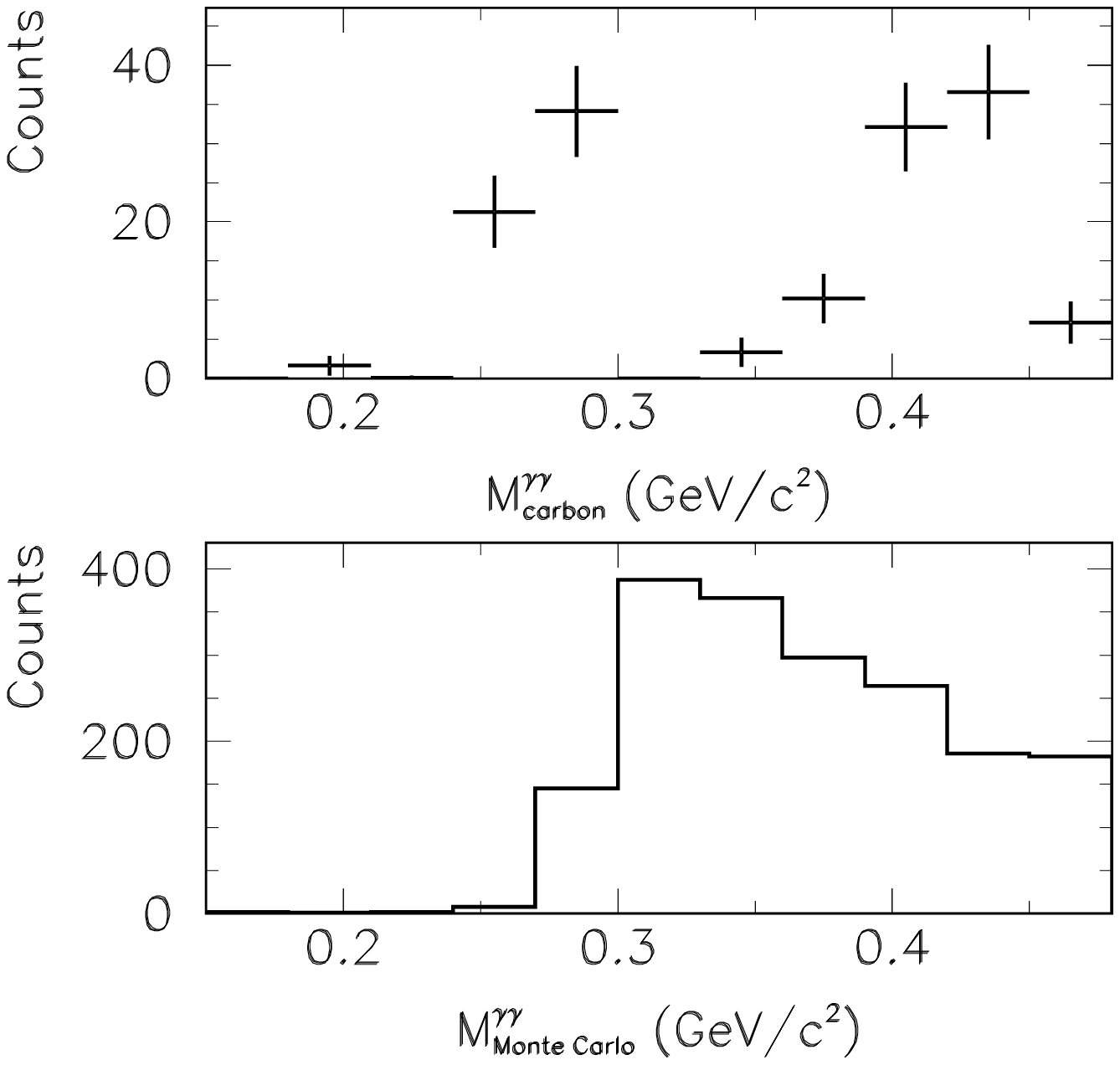,width=0.5\textwidth}}}
\caption{
{\bf Left:} $\gamma \gamma$ invariant-mass spectra obtained for hydrogen and carbon targets. 
The hydrogen spectrum is normalized to carbon in the area around 180~MeV.
{\bf Right (top):} the difference between spectra of the left figure.
{\bf Right (bottom):}  The Monte Carlo results are shown. The $\gamma \gamma$ invariant-mass 
spectrum  for $\sigma \to 2 \gamma$ decay in carbon was generated according to 
Ref.~\cite{chiku}.} 
\label{f02g}
\end{figure}

We obtained the upper limit for BR$(\sigma \to 2\gamma)$ in carbon by comparing 
the $\gamma \gamma$ invariant-mass spectra of the hydrogen and carbon targets. 
Some additional cuts were applied to the spectra shown in Fig.~\ref{fig1} to reduce the 
low invariant mass background. The final spectra obtained for the hydrogen and carbon targets 
are shown in Fig.~\ref{f02g}. The hydrogen spectrum is normalized to carbon in the area next 
to the $\pi^0$ peak (at around 180~MeV). The difference taken between those spectra is compared 
to our Monte Carlo results in Fig.~\ref{f02g}. The Monte Carlo results 
generated according to the prediction by Chiku and Hatsuda~\cite{chiku} were used 
to calculated the CB acceptance. With $148 \pm 48$ $\sigma \to 2\gamma$ event 
candidates obtained, $1.6\times 10^4$ $2\pi^0$ events detected for the carbon target, and
$\approx 15$~\% acceptance, we 
calculate the upper limit for BR$(\sigma \to 2\gamma)$ in carbon to be 
$3.6 \times 10^{-3} \times \phi$ at 90\% C.L., where $\phi$ is the fraction of $2\pi^0$ that is 
produced via the $\sigma$ intermediate state. See the talk by one of us (B.M.K.N.) in these 
proceedings for more details regarding the $\sigma$ contribution.

\section{Summary and Conclusions}

New preliminary measurements of $\pi^0$, $2\pi^0$, and $\eta$ production 
on hydrogen and complex nuclear targets (C, Al, Cu) 
by $\pi^-$ at 750~MeV/$c$ are reported.
\begin{enumerate}
\item The data at 750~MeV/$c$ show that absorption rather than 
medium effects is responsible for the change in the
shape of the $2\pi^0$ invariant mass spectra.
\item The preliminary upper limit, BR$(\sigma \to \gamma \gamma) < 3.6\times 10^{-3} \phi$ at 
90\% C.L., has been obtained for $\sigma \to \gamma \gamma$ on a carbon target at 750~MeV/$c$.
\end{enumerate}

\end{document}